\documentclass[mathleft
]{an}
\usepackage{graphicx}
\usepackage{times}
\begin{document}

\Pagespan{789}{}
\Yearpublication{2008}%
\Yearsubmission{2008}%
\Month{11}%
\Volume{999}%
\Issue{88}%

\title{Probing the kinematics of early-type galaxy halos using
  planetary nebulae}

\author{L. Coccato\inst{1}\fnmsep\thanks{Corresponding author:
  \email{lcoccato@mpe.mpg.de}\newline}
\and  O. Gerhard\inst{1}
\and  M. Arnaboldi\inst{2,3}
\and  P. Das\inst{1}
\and  N. G. Douglas\inst{4}
\and  K. Kuijken\inst{5}
\and  M. R. Merrifield\inst{6}
\and  N. R. Napolitano\inst{7}
\and  E. Noordermeer\inst{6}
\and  A. J. Romanowsky\inst{8}
\and  M. Capaccioli\inst{9,10}
\and  A. Cortesi\inst{6}
\and  F. De Lorenzi\inst{1}
\and  K. C. Freeman\inst{11}
}

\titlerunning{Probing the kinematics of early-type galaxy halos using PNe.}
\authorrunning{L. Coccato et al.}
\institute{Max-Plank-Institut f\"ur Extraterrestrische Physik, Giessenbachstra$\beta$e, D-85741 Garching bei M\"unchen, Germany
\and 
European Southern Observatory, Karl-Schwarzschild-Stra$\beta$e 2, D-85748 Garching bei M\"unchen, Germany
\and 
INAF, Osservatorio Astronomico di Pino Torinese, I-10025 Pino Torinese, Italy
\and
Kapteyn Astronomical Institute, Postbus 800, 9700 AV Groningen, The Netherlands
\and
Leiden Observatory, Leiden University, PO Box 9513, 2300RA Leiden, The Netherlands
\and
School of Physics and Astronomy, University of Nottingham, University Park, Nottingham NG7 2RD, UK
\and
INAF-Observatory of Capodimonte,  Salita Moiariello, 16, 80131, Naples, Italy
\and
UCO/Lick Observatory, University of California, Santa Cruz, CA 95064, USA
\and
Dipartimento di Scienze Fisiche, Universit\'a Federico II, Via Cinthia, 80126, Naples, Italy
\and
INAF - VSTceN,  Salita Moiariello, 16, 80131, Naples, Italy
\and
Research School of Astronomy \& Astrophysics, ANU, Canberra, Australia
}

\received{--}
\accepted{--}
\publonline{later}


\abstract{We present first results of a study of the halo
  kinematics for a sample of early type galaxies using planetary
  nebulae (PNe) as kinematical tracers. 
PNe allow to extend up to several effective radii ($R_e$) the
information from absorption line kinematics (confined to within 1 or 2
$R_e$), providing valuable information and constraints for merger
simulations and galaxy formation models.  We find
that the specific angular momentum per unit mass has a more complex
radial dependence when the halo region is taken into account and that
the halo velocity dispersion is related to the total galaxy
luminosity, isophotal shape, and number of PNe per unit of luminosity
.
 }

\maketitle

\section{Introduction}

Early type galaxies, despite their regular and simple morphology if
compared to the spirals, have a complex dynamical structure which is
mostly dominated by the random motions of the stars.  The dynamical
structure and its relation to the galaxy's structural and morphological
parameters are the relic of galaxy formation and evolution processes.
The connection between morphological properties of early type galaxies
and their kinematics motivated several authors to revise the original
Hubble classification scheme which is influenced by the galaxy's
orientation relative to the observer. Bender \& Kormendy (1996)
distinguish between ``boxy'' and ``disky''; more recently Emsellem et
al. (2007) refined the distinction between ``slow'' and ``fast''
rotators, according to their specific angular momentum per unit mass
instead of the $V/\sigma$ ratio (e.g. Illingworth 1977).
Disky ellipticals have significant rotation with $V/\sigma$ $\geq 1$
and are generally axisymmetric. Boxy ellipticals have little or no
rotation, show a range of values for $V/\sigma$ including strongly
anisotropic systems ($V/\sigma << 1$), can be triaxial, and are generally
more massive (i.e. Bender 1988; Bender et al. 1989; Kormendy \&
Djorgovsky 1989).
Slow rotators appear to be massive systems, are nearly round with a
significant kinematic misalignment, implying a moderate degree of
triaxiality, and span a moderately large range of anisotropies. Fast
rotators appear to be rather flattened systems, with no significant
kinematic misalignments, are nearly axisymmetric, and span a larger range
of anisotropies (Cappellari et al. 2007).

\begin{figure*}
\vbox{
\hbox{
  \includegraphics[width=8.6cm]{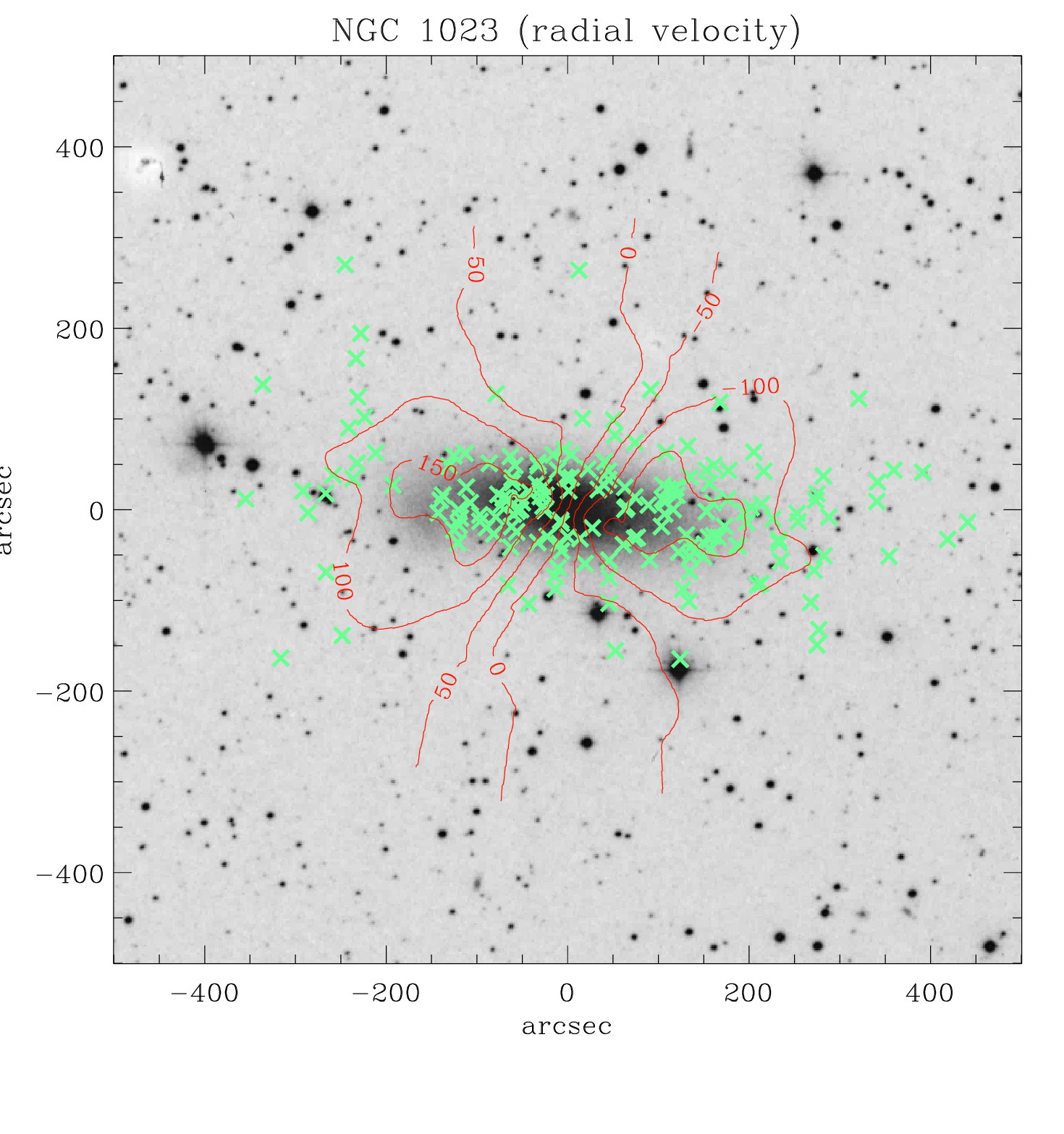}
  \includegraphics[width=8.6cm]{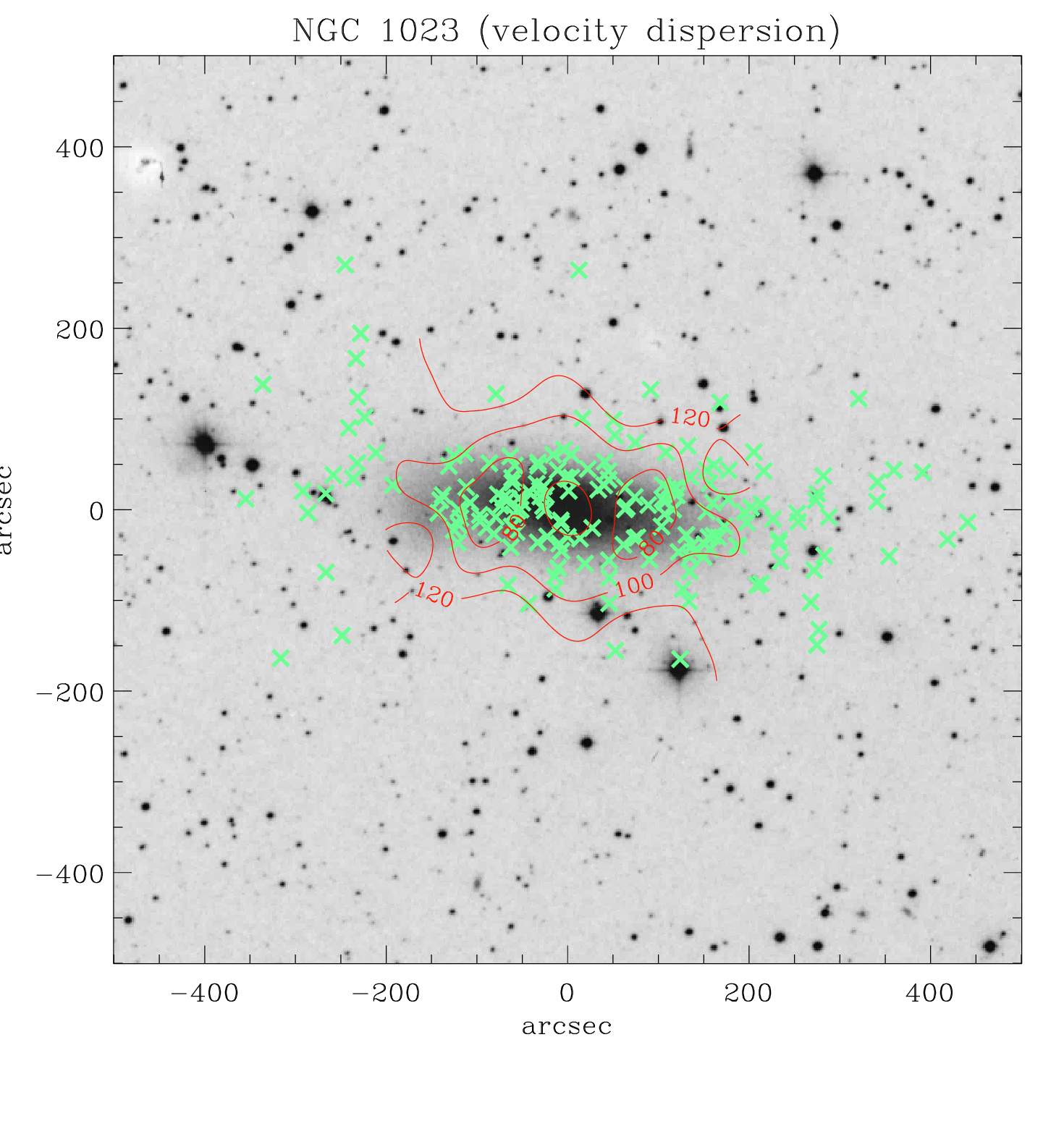}
}
\hbox{
  \includegraphics[width=8.6cm]{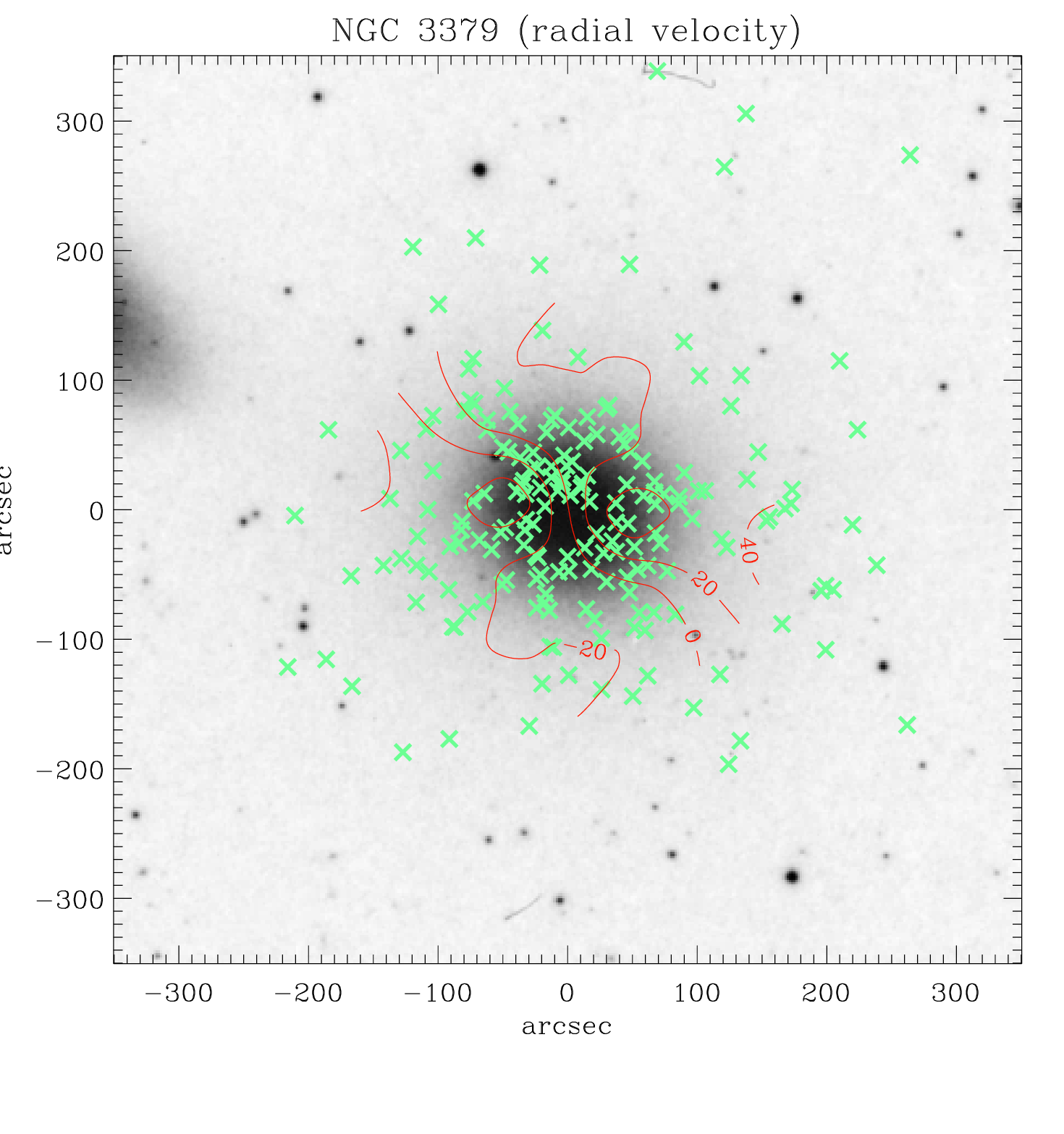}
  \includegraphics[width=8.6cm]{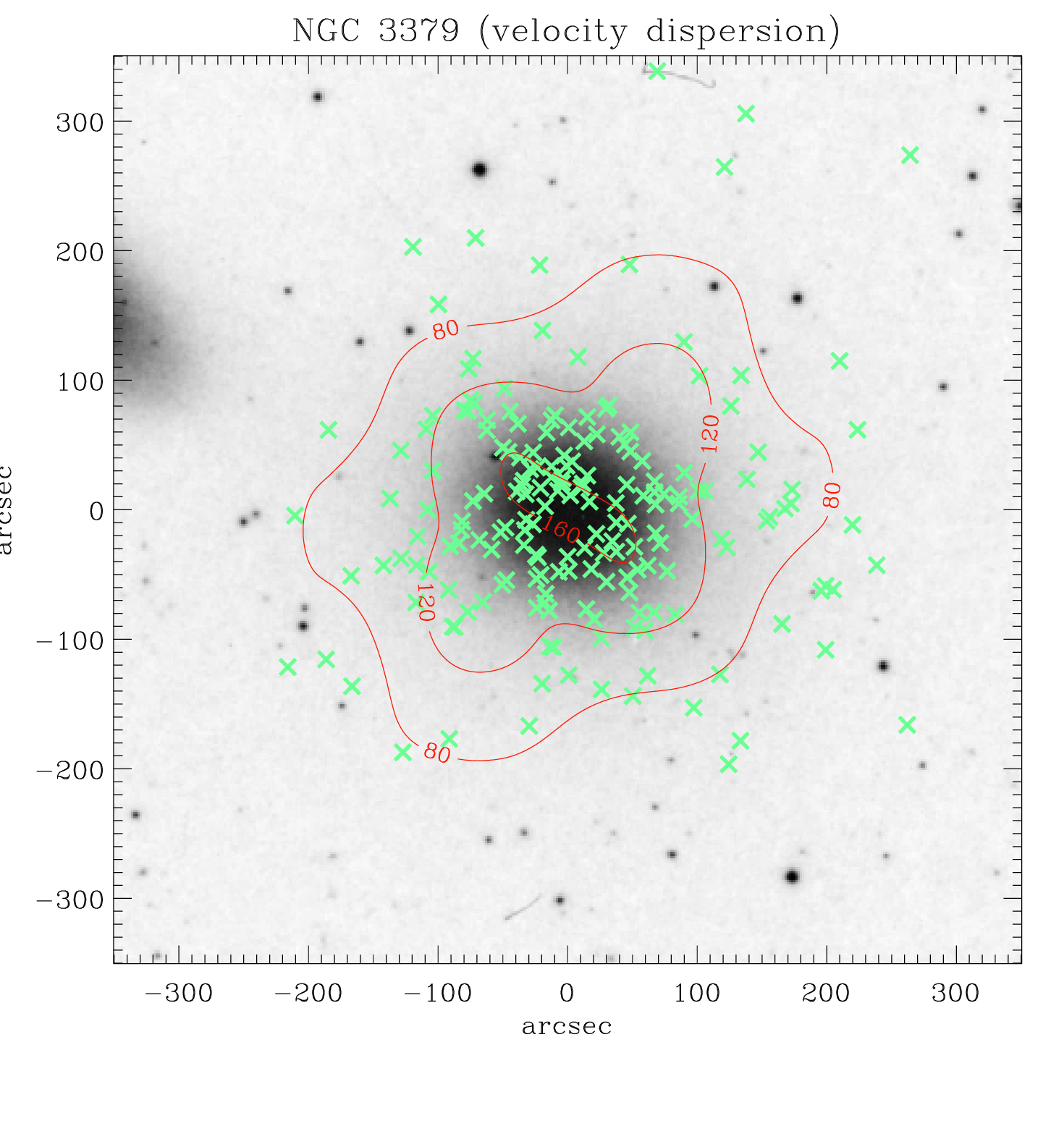}
}
}
\caption{ESO digital sky survey images of NGC 1023 ({\it upper
    panels}) and NGC 3379 ({\it lower panels}) with the PNe positions
  marked with {\it green crosses}. {\it Red lines} represent the
  contour levels of the mean line of sight velocity ({\it left panels}) and
  velocity dispersion ({\it right panels}). 
North is at top, East at left.
}
\label{fig:2d}
\end{figure*}

Differences in galaxy properties can be the result of different
formation scenarios. Numerical simulations of merging galaxies suggest
that the less-luminous fast rotators with disky isophotes are
preferentially formed through a series of minor mergers with less
massive companions. On the other hand, the more luminous slow rotators
with boxy isophotes are thought to form through a violent major
merger between galaxies of similar mass (e.g Naab et al. 1999; Naab \&
Burkert 2003). Moreover, numerical simulations of galaxy formation in
a cosmological context predict particular radial profiles for the
total and dark matter distribution (e.g. Dekel et al. 2005; Naab et
al. 2007) and for kinematic properties such as $V/\sigma$, orbital
distribution and angular momentum (e.g. Abadi et al. 2006, Naab et
al. 2006).
The comparison of merger simulations and model predictions with
kinematic observations out to large radii is a key point for
understanding galaxy mass distributions and formation processes. Spirals
have the advantage that their kinematics can be easily measured from
neutral and ionized gas out to several scale radii, but the situation
is more complex for early type galaxies, in which the absence of gas
and the rapid fall-off of the stellar light allows for kinematic
measurements only within 1 or 2 effective radii (e.g. Bender et
al. 1994; Kronawitter et al. 2000; Saglia et al. 2000).

Planetary nebulae (PNe, hereafter) help to overcome this difficulty,
since their bright O$_{\rm [III]}$ emission makes them easily
observable far from the center of early type galaxies, where the
stellar light is too faint for absorption line spectroscopy (e.g. Hui
et al. 1995; Arnaboldi et al. 1996).
A dedicated instrument was installed at the William Herschel
Telescope: the Planetary Nebulae Spectrograph (PN.S, \linebreak Douglas et
al. 2002). A long-term observational campaign of early type galaxies
has been undertaken, aimed at quantifying the halo kinematics and dark
matter content in these systems.
A series of papers has already been published, presenting the initial
results of the PN.S survey (Romanowsky et al. 2003; Douglas et
al. 2007; Noordermeer et al. 2008; De Lorenzi et al. 2008; Napolitano
et al. 2008). To obtain a general overview of the outer halo
kinematics in these systems and the constraints implied for formation
scenarios, Coccato et al. (2008, C+08 hereafter) studied the kinematic
properties of the outer halos of 16 early-type galaxies using
PNe. Here we present first results from this project.

\section{First results}

We illustrate the kinematic analysis of two galaxies in the project,
NGC 1023 and NGC 3379.  Their PNe samples were taken from Noordermeer
et al. (2008, N+08 hereafter) and Douglas et al. (2007, D+07
hereafter), respectively. Two dimensional velocity and velocity
dispersion fields were obtained applying an adaptive Gaussian kernel
smoothing to the measured PN radial velocities as described in C+08,
and are shown in Figure \ref{fig:2d}.

Small rotation is observed in the E1 galaxy NGC 3379, while higher
rotation velocities are observed in the S0 \linebreak galaxy NGC 1023. The
kinematics of the PNe system is in good agreement with the stellar
absorption line kinematics in the radial region in which both data
sets overlap, as shown in N+08 and D+07. As a general result,
PNe turn out to be reliable tracers of the stars and this property
reflects also in their spatial distribution, not only in their
kinematics. In fact, the radial profile of the PNe number density
agrees well with the stellar surface brightness in the region in
which these data overlap. In the outermost regions, the PNe counts
follow the extrapolation of the fit to the stellar light (see C+08 and
references therein for a more exhaustive compilation).

\begin{figure}[t!]
\includegraphics[width=8.36cm]{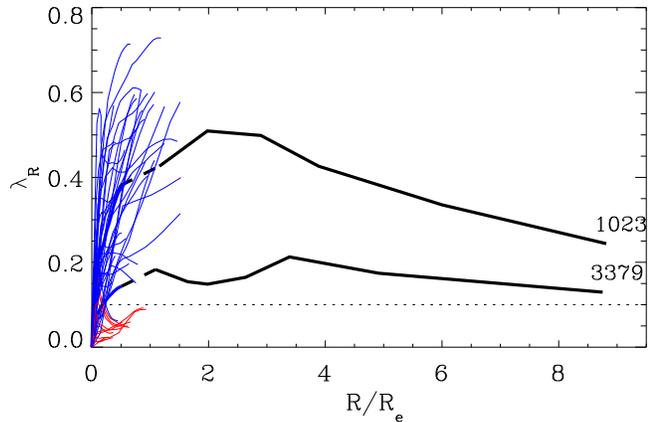}
\caption{{\it Blue} and {\it red solid lines:} $\lambda_R$ profiles
  computed from the SAURON stellar kinematics for fast and slow
  rotators respectively, courtesy of Eric Emsellem. {\it Black solid
    lines:} $\lambda_R$ profiles extracted from the PNe kinematics out
  to large radii. The {\it dotted line} ($\lambda_R = 0.1$) separates
  fast and slow rotator regions.}
\label{fig:lambda}
\end{figure}

The first physical quantity we study using the velocity and velocity
dispersion fields for these 2 galaxies is the radial profile of the
specific angular momentum per unit mass ($\lambda_R$, as defined in
Emsellem et al. 2007). In particular we are interested in probing
$\lambda_R$ into the halo regions, extending the previous information
confined within $0.5R_e$ from integral field absorption line kinematics.
In Figure \ref{fig:lambda} we compare the outer $\lambda_R$ (out to $7-8R_e$,
obtained from PNe kinematics) with the inner values for NGC 1023 and NGC
3379, and with the general trends observed for slow and fast rotators
(Emsellem et al. 2007).
Both NGC 1023 and NGC 3379 are classified as fast rotators, they have
a monotonically rising $\lambda_R$ within $0.5R_e$ reaching 0.4 and
0.15 respectively.  Their halos, however, behave differently. In NGC
3379, $\lambda_R$ remains almost constant, ranging between 0.2 and
0.15; in NGC 1023, $\lambda_R$ keeps rising up to 0.5 at $R=2-3R_e$
and then declines monotonically, almost approaching the slow rotator
regime. This is a consequence of the combined effects of the declining
rotation curve and the rising velocity dispersion profile observed in
NGC 1023 (see N+08).

\begin{figure}[h!]
\includegraphics[width=8.36cm]{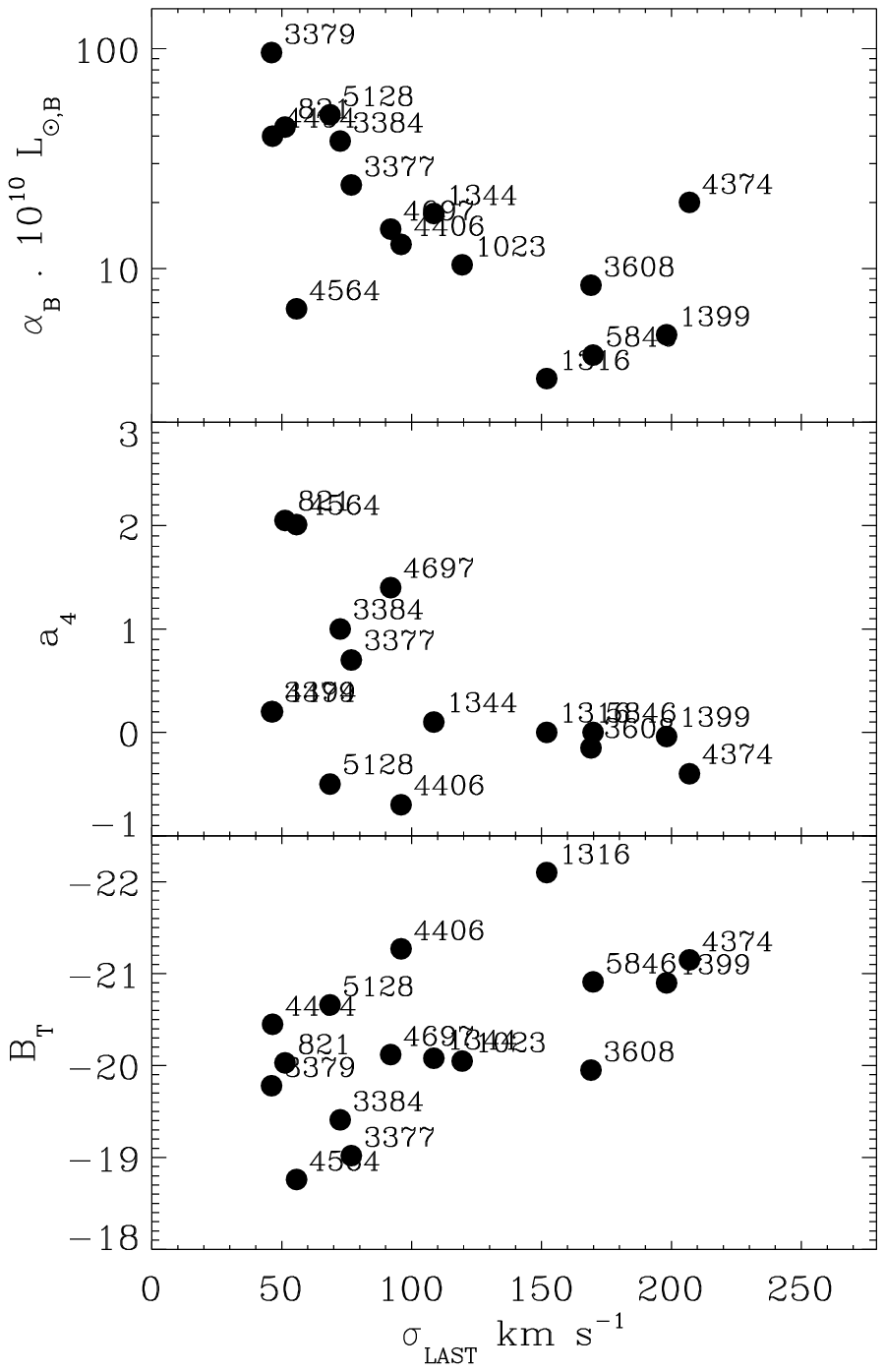}
\caption{Comparison between $\sigma_{\rm LAST}$ and other galaxy
  properties: number of PNe per unit total B-band luminosity ({\it
    upper panel}), isophotal shape ({\it central panel}) and total
  $B$-band magnitude. We refer to C+08 and references therein for the
  computation of these quantities.}
\label{fig:correlations}
\end{figure}

The second aspect of our study is to compare the halo kinematic
properties with other physical properties of the galaxies, such as
luminosity (total $B$-band magnitude), \linebreak shape parameter ($a_4$), and
number of PNe per luminosity in the B band ($\alpha_{B}$, Jacoby
1980). The halo kinematics were parameterized by the outermost value
of the velocity dispersion derived from the PNe data
($\sigma_{\rm LAST}$), and determined for a sample of early type galaxies
from the literature with PNe measurements (see C+08 for references).

In Figure \ref{fig:correlations} we show this comparison with the
following results:

\begin{enumerate}

\item{} More luminous galaxies tend to have larger values of the
  dispersion measured at the outermost observed point. This is related
  to the fact that more massive galaxies have generally higher values of
  the velocity dispersion. 

\item{} Galaxies with higher values of $\sigma_{\rm LAST}$ are preferentially boxy
  in shape ($a_4 < 0$).  This is also a reflection of the known trend
  for massive ellipticals to be more boxy in shape (e.g. Bender et
  al. 1989; Napolitano et al. 2005).

\item{} Galaxies with higher $\sigma_{\rm LAST}$ have smaller $\alpha_{B}$
  values (i.e., less PNe per luminosity in the $B$-band). As discussed in Buzzoni
  et al. (2006), this is probably a consequence of massive early-type
  systems harboring a larger proportion of stars on the Horizontal
  Branch that do not enter the PN stage.

\end{enumerate}

\section{Conclusions}

We have presented first results from the study of the kinematic properties
of early type galaxy halos by means of PNe velocities. Good agreement
is observed between absorption line kinematics and PNe kinematics, as
well as between PNe radial distribution and stellar surface brightness profile
(see C+08 and references therein). This strongly
supports that PNe are reliable tracers of the stellar kinematics.

The results for NGC 1023 and NGC 3379 show that the halo $\lambda_R$
radial profile can show different behavior from the profile measured
within $\sim 1R_e$.  In addition, the PNe kinematics allow us to probe
the relation between the halo
velocity dispersion and other galaxy properties, such as luminosity,
shape parameter and number of PNe per luminosity unit.

The results of this study, even if they are obtained for a small sample of
galaxies, will provide new constraints for models of elliptical galaxy
formation. Most of the merger simulation papers to date compare their
remnants to data within an effective radius or so. We
expect that extending the predictions of the simulations to outer radii
and comparing with data of the kind presented here will shed new light on the
merger formation histories of elliptical galaxies.

\acknowledgements We would like to thank Eric Emsellem for providing
the $\lambda_R$ radial profiles from the SAURON data.

PD is supported by the DFG Cluster of Excellence ``Origin and
Structure of the Universe''.
FDL is supported by the DFG Schwerpunktprogram SPP 1177 ``Witnesses of
Cosmic History''.
MRM is supported by an STFC Senior Fellowship.
NRN has been funded by CORDIS within FP6 through a Marie Curie
European Reintegration Grant, contr. n. MERG-FP6-CT-2005-014774,
co-funded by INAF.
AJR is supported by the National Science Foundation Grant AST-0507729,
and by the FONDAP center for Astrophysics CONICYT 15010003.

\end{document}